# Mechanical Heterogeneity and the Nuclear Mechanome: Towards an Omics-Level Understanding of Nuclear Function

**Lucía Benito-Barca[1], Carlos del Pozo-Rojas[1]\*, Sandra Montalvo-Quirós[1], Ramiro Perezzan Rodríguez[1], Irene Alférez[1], Jorge Barcenilla[1], Diego Herráez-Aguilar[1]\***

[1]*Grupo de Biofísica Computacional, Instituto de Investigaciones Biosanitarias, Universidad Francisco de Vitoria, Carretera Pozuelo-Majadahonda km 1.800, 20223, Pozuelo de Alarcón, Madrid*

*\* Corresponding authors: carlos.delpozo@ufv.es, diego.herraez@ufv.es*

**Abstract**

The cell nucleus is increasingly recognized as a mechanosensitive organelle whose mesoscale mechanical heterogeneity (100 nm–μm) is inseparable from genome regulation yet remains weakly integrated into systems biology and omics frameworks. Here we synthesize experimental and theoretical advances that define a nuclear mechanome: a multidimensional state of mechanical observables spanning lamins, chromatin, nucleoplasm and condensates, and varying across space, time, and cells. We review how intranuclear gradients, chromatin domains and phase-separated bodies generate viscoelastic, poroelastic and plastic niches that shape transcription, replication, DNA repair, epigenetic metabolism, and clonal selection, with particular emphasis on cancer, laminopathies and ageing. We survey invasive, optical and high-throughput mechanophenotyping techniques, together with continuum, shell–polymer, mesoscopic and non-equilibrium models, as a toolkit for mechanomic mapping. Finally, we outline a research agenda towards standardized mechanical descriptors, tissue-level mechano–multi-omic atlases and mechanotherapeutic strategies that explicitly target nuclear mechanotypes as potential biomarkers that can define cellular identity, explain plasticity, and help predict pathological states or the evolution of several diseases.

# 1. Introduction

The eukaryotic nucleus is now understood as a mechanosensitive organelle whose elastic, viscoelastic and poroelastic properties actively influence cell behavior [1–3]. These properties emerge from the coupled mechanics of chromatin, nuclear lamins, the envelope, nucleoplasmic fluid and the perinuclear cytoskeleton, giving rise to distinct deformation regimes in which chromatin dominates small strains while lamins govern large-strain stiffening [4,5]. At mesoscopic scales (100 nm to 10 μm), spatial gradients of stiffness at the lamina–heterochromatin-rich periphery, euchromatin-rich interiors and biomolecular condensates become resolvable and reveal a structured mechanical landscape [3,6–8].

Across these scales, nuclear mechanics is not uniform but varies with cell state, perturbations and lineage [7,9]. Spatial, temporal and intercellular variability arise from chromatin compaction, lamina composition, nucleoplasmic rheology and active processes, each exhibiting measurable heterogeneity with functional consequences [10–14]. This regulated heterogeneity motivates an omics-like perspective: the nuclear mechanome [14], defined as the multidimensional set of mechanical properties, force-transmission pathways and dynamic physical states that characterize the nucleus [14,15]. In this view, mechanical observables (moduli, relaxation spectra, fluctuations, and their distributions) are treated as primary descriptors of nuclear state and as substrates for functional interpretation.

This brief review synthesizes current understanding of mesoscale mechanical heterogeneity and frames it as a functional layer with relevance for genome organization, nuclear metabolism, clonal selection, and disease. We highlight the need to distinguish intrinsic heterogeneity from measurement artefacts, and we outline how mechanome-inspired descriptors could integrate with spatially resolved epigenomic, transcriptomic, proteomic and metabolomic data, with potential applications in cancer, laminopathies and ageing.

## 2. Mechanical foundations of the nucleus

At mesoscopic scales, nuclear architecture operates as a composite mechanical system formed by lamina, chromatin, nucleoplasm and biomolecular condensates (see **Figure 1**). The nuclear periphery constitutes a load-bearing shell in which lamins, the dual membrane and LINC complexes transmit forces across the envelope and couple extracellular and cytoskeletal mechanics to chromatin [1,2]. Gradients in lamin A/C abundance generate spatially patterned stiffness at the nuclear periphery. These patterns shape nuclear morphology, modulate rupture susceptibility by redistributing strain, and influence mechanotransduction by tuning the coupling between the envelope and chromatin [11,15–20].

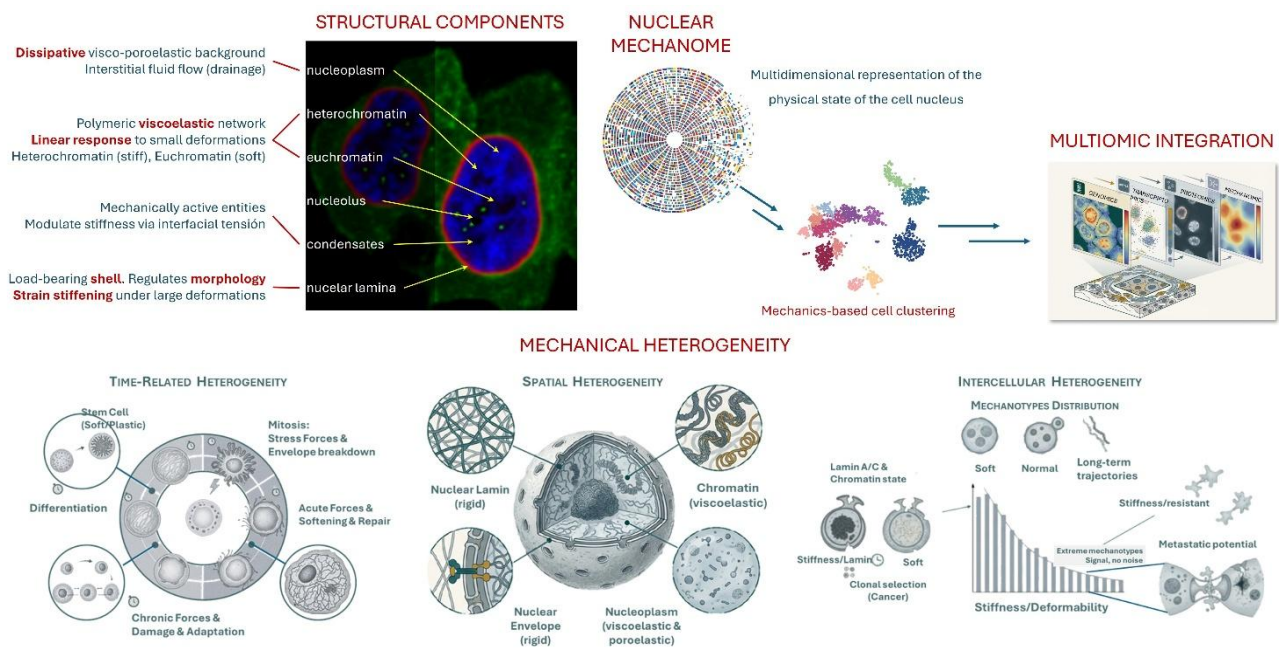

**Figure 1. The nuclear mechanome as a heterogeneous and integrative physical layer of nuclear function.** The nucleus is shown as a mechanically composite organelle whose physical state emerges from the coupled contributions of the nuclear lamina, chromatin, nucleoplasm, and biomolecular condensates (left image). The lamina acts as a load-bearing shell that regulates nuclear shape, strain stiffening, and rupture susceptibility, while chromatin forms a viscoelastic polymer network in which heterochromatin is stiffer than euchromatin and dominates the small-strain response. The nucleoplasm provides a dissipative visco-poroelastic background, and condensates function as mechanically active domains that locally modulate stiffness. Together, these components generate structured mechanical heterogeneity at the mesoscale, which can be represented as a multidimensional nuclear mechanome and integrated with other omics layers within a multi-omic framework.

Within this shell, chromatin behaves as a viscoelastic polymer network whose compaction state primarily contributes to small-strain mechanical response. Heterochromatin stiffens the nucleus and promotes failure when lamins are limiting, whereas more open euchromatin supports compliant behavior [5,21,22]. Chromatin organization (peripheral heterochromatin, interior euchromatin and lamina-associated domains) imposes mesoscale stiffness gradients and couples three-dimensional genome architecture to boundary mechanics [19,23]. Compartmentalization into TADs and larger chromatin domains further introduces mechanically distinct phases associated with differential densities and dynamic constraints [24–31].

The nucleoplasm adds a dissipative visco-poroelastic background. AFM indentation, microrheology and needle penetration reveal multiple relaxation times originating from chromatin rearrangements, viscous drag and interstitial fluid flow [7,12,32]. Rapid loading increases apparent stiffness by fluid trapping, whereas slower deformations permit drainage and stress redistribution, behavior captured by poroelastic modelling [3,33,34]. Mesoscale simulations treating chromatin and nucleoplasm as interacting liquid-like phases reproduce this emergent viscoelasticity and domain reorganization [35–37]. Condensates such as nucleoli, speckles and HP1α-rich droplets act as mechanically active entities whose interfacial tension and crosslinking modulate chromatin stiffness and internal force propagation [10,30,31,37–39].

This composite architecture yields three principal mechanical regimes: (i) chromatin-dominated linear elasticity at small strains, (ii) lamin-dependent strain stiffening at larger deformations, and (iii) poroelastic rate dependence governed by nucleoplasmic permeability. These regimes define core coordinates of the nuclear mechanome, shaping how stiffness, dissipation and mechanical adaptability vary across space, time, and cell populations.

## 3. Mechanical heterogeneity at the mesoscale

Mesoscale nuclear mechanics resolves into a structured and polarized landscape rather than a uniform material state [10]. At spatial scales of 100 nm to several micrometers, a stiff, lamin- and heterochromatin-enriched cortex surrounds a more compliant interior [40]. Lamina-associated domains tether chromatin to the envelope and concentrate load at the periphery, producing radial segregation of dense chromatin and softer internal regions, a pattern reproduced by mesoscale liquid-like simulations [17,20,36,37,39]. Experimental mapping by AFM microrheology, multiple-particle tracking and Brillouin-based imaging consistently reveals higher moduli, slower relaxation and reduced mobility near the nuclear surface relative to central euchromatic zones and nucleoli [7–9,41–43]. At finer scales, phase-separated condensates, A/B compartments and TAD-level organizations form mechanically distinct phases whose viscoelasticity and interfacial properties evolve with epigenetic and transcriptional state [22,25–31,37–39]. These spatial patterns define local mechanotypes and establish the first axis of the nuclear mechanome.

Nuclear mechanics also varies dynamically over time as cells traverse the cell cycle or respond to mechanical and genotoxic stress. During G1–S–G2, replication, chromatin decondensation–recondensation cycles and lamin phosphorylation subtly adjust nuclear rheology [1,2,44]. Approaching mitosis, increased nuclear tension, envelope–chromatin coupling and controlled lamin disassembly convert a poroelastic, envelope-stabilized structure into a condensed, chromatin-dominated body whose mechanical state is tightly linked to checkpoint regulation [46–48]. External forces add further trajectories: cyclic stretching, compression, shear and confined migration induce heterochromatin loss, softening and adaptive reorientation, or trigger ATM-dependent decondensation following DNA damage, whereas nuclei that fail to decondense remain stiff and rupture-prone, accumulating secondary breaks [47,49]. Over longer timescales, differentiation increases rigidity through lamin reinforcement and heterochromatin accumulation, while senescence and ageing produce enlarged, irregular nuclei with defective lamins, altered constitutive heterochromatin and heightened fragility [10,35,50–52]. These temporal trajectories populate distinct regions of nuclear mechanical state space.

Superimposed on spatial and temporal variability is substantial intercellular heterogeneity. Even genetically homogeneous populations display broad distributions of lamin A/C abundance, chromatin compaction and nuclear geometry [4,11,17,29,49]. These parameters define partially independent axes of stiffness, viscoelasticity and rupture susceptibility, shaping access to transcriptional and repair machinery [1,21]. Cancer exemplifies this heterogeneity: subclones differ in lamina composition, heterochromatin organization and nuclear size, generating mechanotypes associated with migratory capacity, drug resistance and metastatic competence, and remodeled during repeated confined migration or shear exposure [2,2,53–58]. High-throughput deformability measurements reveal that extreme mechanical tails, rather than mean values, correlate with invasion, treatment response and disease stage [59].

Together, spatial, temporal, and intercellular variability define the three principal axes of mechanical heterogeneity. These axes constitute the structured state space of the nuclear mechanome and underlie its functional relevance for genome regulation, cellular adaptation, and disease.

## 4. The nuclear mechanome as an omics layer

Mechanomics was proposed as a systems-level discipline that catalogues how mechanical cues shape biological function, in analogy to genomics or proteomics [15,60–62]. In this framework, the nuclear mechanome denotes the complete set of mechanical parameters, force-transmission pathways and mechanoresponsive outputs that define the nuclear state (see **Figure 1**). Operationally, it includes structural descriptors (lamin isoform abundance, chromatin compaction, nuclear size and shape), material properties (elastic moduli, viscosities, relaxation times, strain-stiffening coefficients, poroelastic permeabilities) and dynamical observables (fluctuation spectra, active noise, energy dissipation), together with mechanotransductive outputs such as chromatin accessibility and transcriptional or epigenetic responses [3,3,12,63]. Each nucleus occupies a point in this mechanical state space, while populations trace distributions and trajectories over space, time, and cell identity.

**Table 1** shows the key mechanical and dynamic properties for characterizing the nuclear mechanome and its correspondence with instrumental techniques.

Mechanomics was conceived from the outset as an omics layer to be read jointly with genomic, transcriptomic, and proteomic data. In the nucleus, the coupling is direct: forces transmitted via adhesions, cytoskeleton and nuclear envelope remodel chromatin conformation, epigenetic marks and transcriptional programs, embedding mechanical information in classical omics readouts [21,64]. Mechanical state behaves as an epigenetic regulator that modulates accessibility, histone modification landscapes and higher-order genome organization [24,29,65]. Strain mapping, deformability cytometry and related platforms, combined with transcriptomics, now generate mechanotranscriptomic datasets and mechanical feature matrices analogous to expression matrices [59,61,66–68]. Spatial mechanotranscriptomics extends this logic to tissues by co-registering mechanical context with in situ gene expression [67,69].

Within this framework, variability in nuclear mechanics becomes a structured source of functional diversity. Single-cell studies show that differences in stiffness, viscoelasticity and lamin or chromatin composition influence differentiation trajectories and gene expression [11,14,50,70]. Heterochromatin-driven softening under stress protects against DNA damage, indicating that temporal modulation of mechanics is an adaptive program [49]. Distinct chromatin domains, lamin microdomains and condensates can be viewed as mechanical niches that condition local biochemical reactions [4,10]. In cancer, altered distributions of nuclear mechanotypes contribute to malignant progression and offer diagnostic leverage; not only mean stiffness, but also the shape and tails of mechanical distributions correlate with invasion and treatment response [21,42,55,59,66,67].

Finally, the mechanome concept naturally invites a feedback-loop view in which mechanics, 3D genome organization and gene regulation are tightly coupled. Forces integrated by lamina, LINC complexes and chromatin reorganize folding, compartmentalization and nuclear bodies, altering regulatory contacts [29,71–73]. In turn, epigenetic and transcriptional programs tune lamins, chromatin compaction and condensates, reshaping nuclear mechanics and active fluctuations [10,24,65,74]. In this sense, mechanomic descriptors function as an additional coordinate system for interpreting and integrating multi-omics datasets.

## 5. Measuring the nuclear mechanome

Quantifying the nuclear mechanome requires instruments that sample complementary regions of mechanical state space (see **Figure 2**). Low-throughput, high-content techniques directly deform nuclei with controlled forces and resolve local mechanics. AFM indentation, needle-based AFM, optical tweezers and force mapping on adherent cells or isolated nuclei provide force–displacement curves from which effective moduli, relaxation times and poroelastic parameters are inferred, including envelope and lamina failure thresholds linked to lamin composition and chromatin compaction [7,7,33,34,43,75]. Micropipette aspiration, microharpoon assays and envelope-tethered tweezers probe large-strain rheology and nucleus–cytoskeleton anchoring with tight control of geometry and loading history [2,7,75].

High-throughput, population-scale approaches trade directness for statistics and physiological relevance. Brillouin microscopy yields label-free 3D maps of longitudinal modulus across periphery, interior and nucleoli, especially when cross-calibrated with AFM and fluorescence [7,8,44]. Real-time deformability cytometry and related microfluidic constriction assays hydrodynamically deform $10^4$–$10^6$ cells and extract effective stiffness, viscoelastic proxies and rupture probabilities, revealing population-level distributions and tails of mechanotypes [59,68]. Multiple-particle tracking of endogenous granules or tracers reports local mean squared displacements and anomalous diffusion exponents that can be inverted into frequency-dependent storage and loss moduli, resolving intranuclear microrheology [42,76].High-throughput, population-scale approaches trade directness for statistics and physiological relevance. Brillouin microscopy yields label-free 3D maps of longitudinal modulus across periphery, interior and nucleoli, especially when cross-calibrated with AFM and fluorescence [7,8,44]. Real-time deformability cytometry and related microfluidic constriction assays hydrodynamically deform $10^4$–$10^6$ cells and extract effective stiffness, viscoelastic proxies and rupture probabilities, revealing population-level distributions and tails of mechanotypes [59,68,77]. Multiple-particle tracking of endogenous granules or tracers reports local mean squared displacements and anomalous diffusion exponents that can be inverted into frequency-dependent storage and loss moduli, resolving intranuclear microrheology [42,76].

Correlative imaging integrates these mechanical readouts with genome architecture. Super-resolution microscopy and Hi-C (ref) define chromatin domains, lamina-associated regions and TADs, while traction force microscopy, AFM and deformation microscopy relate local density and anchoring to regional stiffness and strain localization [25,28,36,36,37,78,79]. Mechanics-indexed single-cell and spatial designs couple RT-DC or optical mapping with RNA-seq, ATAC-seq or spatial transcriptomics, turning mechanotypes into explicit features in omics matrices [61,66–69].Correlative imaging integrates these mechanical readouts with genome architecture. Super-resolution microscopy and Hi-C define chromatin domains, lamina-associated regions and TADs, while traction force microscopy, AFM and deformation microscopy relate local density and anchoring to regional stiffness and strain localization [25,28,36,36,37,78,79]. Mechanics-indexed single-cell and spatial designs couple RT-DC or optical mapping with RNA-seq, ATAC-seq or spatial transcriptomics, turning mechanotypes into explicit features in omics matrices [61,66–69].

Interpreting these data demands treating all observables as distributions (variance, skewness, and tails) not just means and carefully separating biological heterogeneity from artefacts. Nuclear mechanics depends on cell cycle, size, nucleus–cytoplasmic ratio, adhesion geometry, and matrix stiffness, while probe geometry, indentation depth, optical path, or flow alignment can bias measurements. Recent reviews emphasize the need for benchmarks, reporting standards, multimodal cross-validation and open analysis pipelines as prerequisites for a robust mechanomic description [3,75].

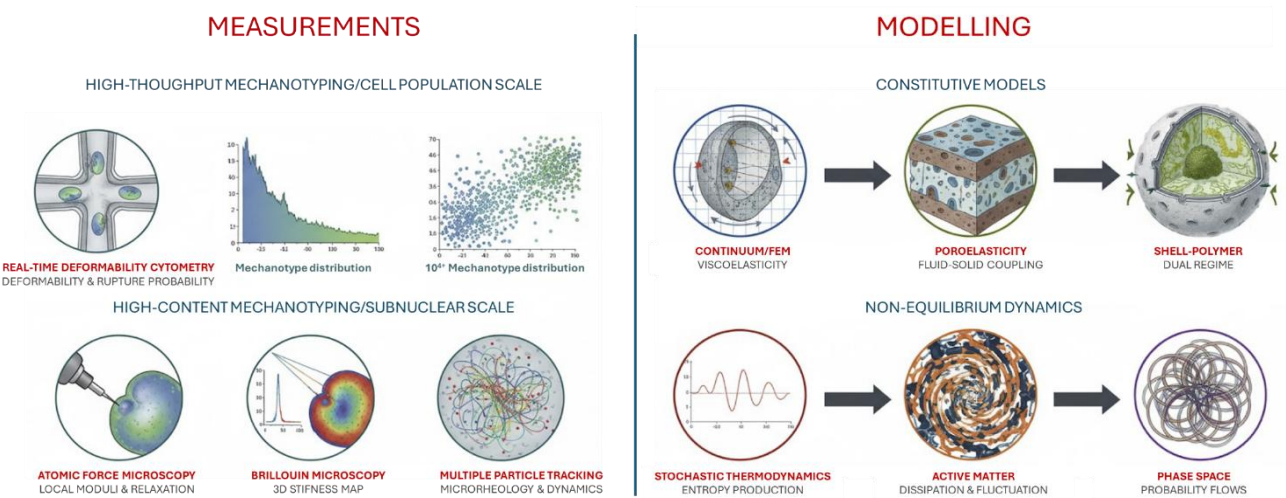

**Figure 2 Experimental measurement and physical modelling of the nuclear mechanome.** The nuclear mechanome can be accessed through complementary experimental and theoretical approaches. On the measurement side, high-throughput mechanophenotyping techniques, such as real-time deformability cytometry, capture population-level distributions of nuclear deformability and rupture probability, while high-content methods including atomic force microscopy, Brillouin microscopy, and particle-tracking microrheology resolve local stiffness, relaxation, and dynamic fluctuations. On the modelling side, constitutive frameworks (ranging from continuum and finite-element descriptions to poroelastic and shell–polymer models) link observed deformations to underlying material properties. These approaches are extended by non-equilibrium and stochastic descriptions that account for active forces, dissipation, and fluctuations, enabling a multidimensional, physically grounded representation of nuclear mechanical states.

**Table 2** summarizes the principal techniques used to extract the biomechanical observables that characterize the mechanical state and response of cell nuclei.

# 6. Physical models and non-equilibrium dynamics

Within a mechanomic framework, physical models compress heterogeneous measurements into interpretable parameter sets (see **Figure 2**). Continuum and finite-element formulations treat the nucleus as a mechanically heterogeneous body in which lamina and chromatin or nucleoplasm obey distinct constitutive laws and are coupled through realistic geometries [1]. Linear and non-linear viscoelastic models reproduce anisotropic shape changes and relaxation under AFM, micropipette or microfluidic loading and yield lamina and nucleoplasmic moduli in the kPa–sub-kPa range. Poroelastic extensions model the chromatin–lamina scaffold as a porous solid saturated with nucleoplasmic fluid, explaining rate-dependent stiffness and drainage times [3,34,80]. Hybrid shell–polymer models represent the nucleus as a thin elastic shell surrounding a viscoelastic polymer interior and account for chromatin-dominated small-strain response, lamin A/C–dependent strain stiffening and buckling suppression [5,23,63,81].

At the mesoscale, soft-matter and polymer approaches describe chromatin as interacting liquid-like phases or coarse-grained chains whose surface tension, viscosity and interaction strengths control domain segregation, radial positioning and emergent stiffness [36–39,82]. These models are particularly useful for translating domain patterns and strain maps into effective mechanical parameters. Validation remains a central challenge. Hobson and colleagues emphasize constraining models with multiple independent measurements: micromanipulation, Brillouin microscopy, deformation microscopy, super-resolution imaging and Hi-C—to avoid parameter degeneracy and to test mesoscale predictions [3,8,9,24,75,83].

A summary of the different types of models and simulations used to describe nuclear mechanics can be found in **Table 3**.

Beyond static parameters, the nuclear mechanome is maintained by continuous energy consumption. Chromatin loci and nuclear bodies display nonthermal fluctuations, anomalous diffusion and superdiffusive bursts driven by ATP-dependent remodeling and cytoskeletal forces [42,43,63]. Isolated nuclei actively adapt to imposed forces and trigger mechanotransduction, behavior that cannot be captured by equilibrium elasticity [84,85]. Stochastic thermodynamics offers a language to quantify this activity: broken detailed balance, probability currents and entropy production provide additional coordinates of mechanomic state [86–91]. Phenomena such as heterochromatin-driven softening under stress or maintenance of specific 3D architectures can be viewed as dissipative strategies with energetic costs and benefits [10,49,88]. Robust mechanotypes emerge when feedback between mechanics, chromatin state and gene regulation stabilizes attractor-like regions in this active landscape, whereas plasticity and mechanopathology reflect noise-enabled transitions, particularly in development, cancer and laminopathies [20,63].

## 7. From mechanics to biological function

**Mechano-dependent regulation of nuclear metabolism.** The nuclear mechanome shapes genome accessibility, transcription, replication, and repair. Forces transmitted to the nucleus alter chromatin compaction, nucleosome spacing and higher-order folding, modulating access to regulatory elements and transcriptional machinery [24,65]. Compressive and tensile stresses redistribute heterochromatin and euchromatin, changing local stiffness and the energetic cost of domain opening [4,29]. Direct chromatin stretching via integrin-mediated forces can acutely upregulate transcription at specific loci, establishing a causal link between tension and gene activation [72]. Increased heterochromatin compaction stiffens nuclei, reduces transcriptional output and protects the genome from mechanically induced damage [30,49,52].

Replication and repair operate within this heterogeneous rheological background. During S phase, replication factories traverse a viscoelastic chromatin network whose rearrangements on multiple timescales are required for efficient fork progression [65,83]. Abnormal stiffening or tethering promotes replication stress, while extreme deformations during confined migration stretch or rupture chromatin, coupling replication to DNA damage [77,92]. DNA lesions activate ATM signaling, which drives global chromatin decondensation, H3K9me3 loss and nuclear softening; nuclei that fail to decondense remain rigid and rupture-prone, accumulating secondary breaks [48]. Mechanically driven heterochromatin redistribution near breaks can bias pathway choice between homologous recombination and non-homologous end joining [47]. Forces also regulate epigenetic metabolism by controlling access of modifying enzymes and remodelers, while condensates such as nucleoli and speckles act as mechanically sensitive reaction compartments that partition biochemical fluxes [31,32,38,40,93,94].

**Nuclear mechanosignaling and control of identity**. Nuclear mechanosignaling relies on a continuous axis from extracellular matrix to chromatin. Integrin-based adhesions transmit matrix forces to the cytoskeleton, which couples to the envelope via nesprin–SUN LINC complexes, lamins and lamina-associated domains, embedding genes in distinct periphery–interior mechanical niches [1,17,31,64]. Perinuclear actin, vimentin and microtubules shape how loads are routed and how rupture is prevented [2,92]. LINC complexes, lamins and nuclear pore complexes act as mechanosensors: they integrate strain over time, modulate transport of factors such as cyclin B1 and influence pathways including SRF/MAL and YAP/TAZ [2,45,46,64,84]. In parallel, chromatin stretching and reorganization under load reshape promoter–enhancer contacts and transcriptional bursting [21,72].

These pathways steer trajectories in a coupled mechanical–epigenetic landscape. Lamin A levels scale with tissue stiffness, and adaptation of nuclear mechanics to matrix stiffness promotes lineage commitment in mesenchymal stem cells [11]. Soft, deformable nuclei in naïve stem cells stiffen during differentiation as lamins and chromatin reorganize, stabilizing lineage-specific transcriptional programs [50]. Mechanical heterogeneity during differentiation correlates with gene expression variability and exploration of alternative fates [70]. Changes in matrix stiffness, topography or tension can reprogram identity by reshaping nuclear shape, lamina architecture and chromatin organization, a process exploited during development and subverted in cancer [21,92,94]. In mechanomic terms, cell identity corresponds to attractor-like regions in non-equilibrium mechanical–epigenetic space.

**Mechanical heterogeneity, clonal selection, and pathology**. Cancer illustrates how mechanical heterogeneity drives selection. Tumor cells repeatedly deform their nuclei during invasion, intravasation and extravasation. Softer nuclei traverse constrictions more readily but suffer frequent envelope rupture and DNA damage, whereas stiffer nuclei resist deformation but may be excluded mechanically [19,77]. Nuclear mechanopathology (altered lamins, chromatin, morphology) both reflects and reinforces malignant programs [92]. Subclonal variation in lamin A/C, heterochromatin content and nuclear size broadens mechanotype distributions, and high-throughput mechanophenotyping reveals skewed deformability distributions in cancer, where rare extreme mechanotypes have been related to metastasis or failure [42,54,59,68].

Laminopathies and ageing provide complementary examples of mechanome mistuning without malignancy. LMNA mutations and associated defects cause tissue-specific dystrophies and progeroid syndromes characterized by altered stiffness, increased fragility and defective mechanotransduction [2,17,54]. Ageing nuclei enlarge, lose laminar integrity and re-pattern constitutive heterochromatin, leading to chronic DNA damage and mechanical decline [35,47,52]. Accumulation of prelamin A or progerin further stiffens envelopes and promotes blebbing [2,17]. In these settings, the nuclear mechanome drifts away from tissue-appropriate attractors, compromising repair, adaptive mechanosignaling and long-term function [94].

# 8. Perspectives and multi-omic futures

The natural outcome of the nuclear mechanome concept would be an atlas: a 3D (or time-dependent 4D) "mechanomic map" that co-registers nuclear mechanical observables with molecular state in intact tissues. Brillouin microscopy, deformation microscopy and related label-free modalities already resolve structured contrasts in and around nuclei in thick samples, capturing local variations in longitudinal modulus and strain fields in vivo [8,9,44,90]. High-throughput platforms such as RT-DC, ELASTomics and microfluidic deformability assays embed effective stiffness and deformability indices into single-cell datasets, revealing skewed distributions of mechanotypes [3,8,9,59,61,66–68]. Spatial mechano-transcriptomics extends this logic to tissues by jointly mapping gene expression and mechanical context [69]. A mechanomic atlas would consist of co-registered fields of stiffness, relaxation, hydraulic response and rupture risk layered with transcriptomic, epigenomic and metabolomic information, analyzed with joint embeddings where mechanical descriptors enter as a full omics layer [9,86,88,90,95].

Several gaps still limit this vision. Nuclear energy dissipation and entropy production remain rarely measured, despite available estimators from stochastic thermodynamics [3,12,83,86–90]. Brillouin contrast is still debated and requires systematic cross-calibration before being treated as a quantitative modulus [8,12,44]. Continuum, poro-viscoelastic, polymer and phase-field models remain under-validated and only partially linked to Hi-C, super-resolution imaging and in vivo deformation fields [37,75,82,90,96]. Mechanomic knowledge is also biased toward cell lines; high-resolution nuclear mechanics co-registered with spatial omics in primary human tissues is still rare [8,44,54,92].

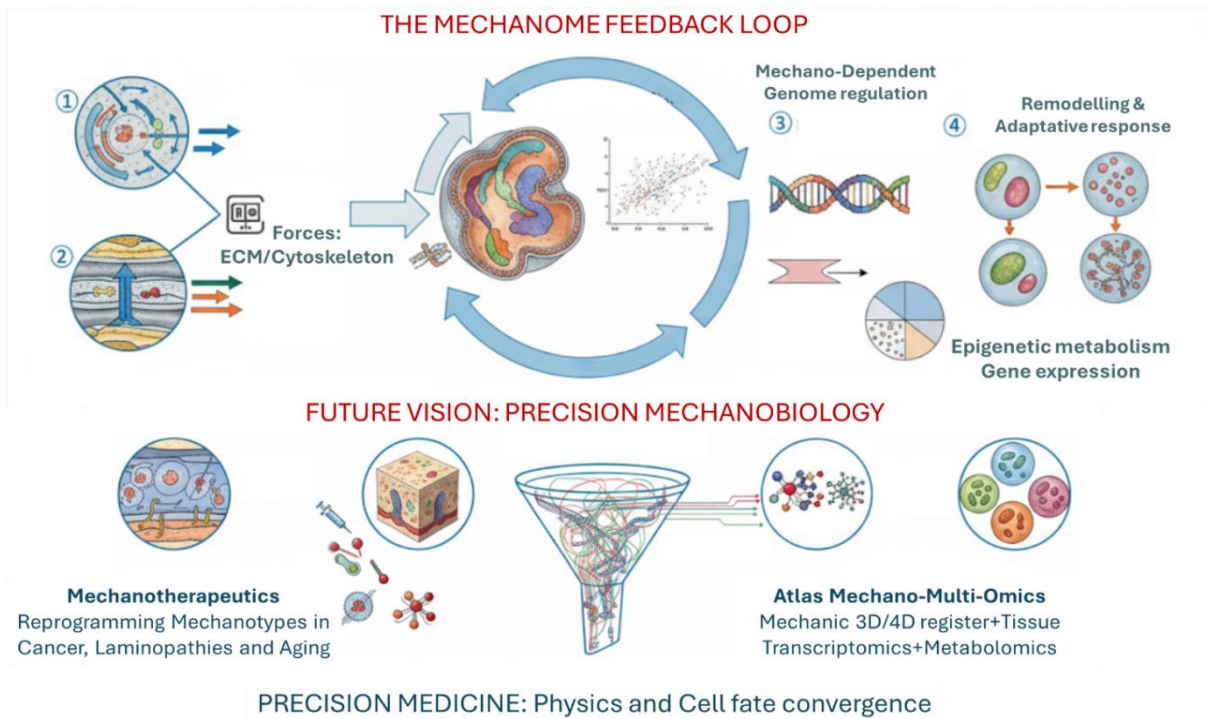

**Figure 3 Feedback between nuclear mechanics, genome regulation and cell fate, and its implications for precision mechanobiology.** Mechanical cues transmitted from the extracellular matrix and cytoskeleton converge on the nucleus, shaping its mechanical state through changes in lamins, chromatin organization and nucleoplasmic dynamics. This nuclear mechanome state feeds back onto genome regulation by modulating chromatin accessibility, epigenetic metabolism, and gene expression, thereby influencing cell phenotype and adaptive responses. Reciprocal remodeling of nuclear mechanics closes a feedback loop that links mechanical history to functional outcomes. At the organismal scale, this framework motivates a future vision of precision mechanobiology, in which targeted mechanical perturbations and mechanotype reprogramming, combined with atlas-level mechano–multi-omic integration, enable refined stratification and intervention in cancer, laminopathies and ageing.

As depicted in **Figure 3,** emerging lines of research point to mechanomics as both readout and handle for precision mechanobiology. Targeted nuclear perturbations using microfabricated or opto-/magnetomechanical tools, combined with time-resolved multi-omics, will move from correlation to causality [12,38,84,86,88]. Clinical-scale panels of mechanomic descriptors derived from RT-DC, microfluidics or optical methods could complement genomic and transcriptomic biomarkers for patient stratification [8,44,61,66–68]. Mechanotherapeutic strategies targeting the ECM–cytoskeleton–nucleus axis and nuclear-intrinsic regulators (lamins, heterochromatin, condensate components) are beginning to be explored in cancer, laminopathies and fibrotic or aged tissues [2,17,32,35,47,52,54,64,65,92–95].

## Conclusions

The nucleus is not a passive container for the genome, but a mechanically heterogeneous, actively driven organelle whose physical state is inseparable from function. At mesoscopic scales, lamins, chromatin, nucleoplasm and condensates form a coupled mechanical system in which periphery–interior gradients, domains and nuclear bodies create a landscape of mechanical niches. Viscoelastic, poroelastic and plastic behaviors are best understood as useful descriptive coordinates in a nuclear mechanome: a multidimensional, time-dependent state space that encodes how the genome senses, filters and amplifies forces, and how these forces reshape genome organization, nuclear metabolism, and cell fate.

Framing nuclear mechanics as an omics layer places it alongside genomics, epigenomics, transcriptomics and proteomics. Within this framework, mechanical heterogeneity becomes a structured and predictive source of functional diversity rather than experimental noise. Distributions of mechanotypes, including their extreme tails, bias clonal selection, support plasticity and, when distorted, contribute to cancer, laminopathies and ageing.

Looking forward, the convergence of advanced mechanical measurements, spatial multi-omics and non-equilibrium theory enables atlas-level descriptions of the nuclear mechanome in human tissues and positions nuclear mechanics as both readout and handle for precision mechanobiology and mechanotherapy.

## Declaration of use of artificial intelligence

Generative artificial intelligence tools were used in a limited and transparent manner during the preparation of this manuscript. Specifically, AI-based tools were employed for the generation of some conceptual illustrations and for grammatical and stylistic language correction. In addition, AI-based search engines were used to support bibliographic searches and literature exploration. All scientific content, interpretation, and conclusions were defined, verified, and approved by the authors, who take full responsibility for the accuracy and integrity of the work.

## Research funding

This work was funded by the Francisco de Vitoria University (UFV2025-55 - RBC-UFV II and UFV2025-48 - GAIA-UFV II) and the Spanish Ministry of Science (LEUKODOMICS Project TED2021-132296A-C53). Irene Alferez gratefully acknowledges the funding from the Programa de Jovenes Investigadores 2025 of Comunidad de Madrid (Spain). Jorge Barcenilla gratefully acknowledges the funding from the Programa de Jovenes Investigadores 2025 of Comunidad de Madrid (Spain). Lucia Benito gratefully acknowledges the IFP predoctoral grant from the Francisco de Vitoria University (IFP2023).

## Acknowledgements

The authors wish to thank Prof. F. Monroy for enlightening conversations and insightful comments.

**Table 1 Measuring the nuclear mechanome: key readouts and instrumental techniques associated.** The table shows the key mechanical and dynamic properties for characterizing the nuclear mechanome and its correspondence with instrumental techniques. The main properties of the cell nucleus are summarized, indicating their typical units and ranges, the spatial and temporal scales at which they are defined, and the experimental techniques used for their measurement. An estimate of experimental throughput and the most natural integration frameworks with omics approaches are also included, highlighting how different mechanical descriptors of the nucleus can be linked with transcriptomics, epigenomics, proteomics, and metabolomics to construct an integrated view of mechanogenomic regulation.

| Property | Unit & range | Spatial scale | Temporary scale | Instrumental Techniques | Throughput | Most natural omics integration |
|---|---|---|---|---|---|---|
| **Apparent elastic modulus**: small-strain stiffness from force–displacement) | kPa ($\approx$0.1–10 kPa for nuclei; soft stem cell nuclei at low end, lamin A/C–rich at high end) | Submicron to whole nucleus | ms to s | AFM indentation / mapping, micropipette aspiration, micro harpoon, optical/magnetic tweezers (surface-tethered) | Low–medium ($10^1$–$10^3$) | Single-cell RNA-seq and ATAC-seq on mechanically profiled cells; integration with lamin and chromatin-mark proteomics |
| **Viscoelastic relaxation times / spectra**: time-dependent stress relaxation, creep, frequency response | Relaxation times (s); frequency-dependent G,' G'' (Pa) | Local patches or whole nucleus | ms to s | AFM stress-relaxation / oscillatory modes; micropipette aspiration; bulk or local micro rheology; tweezer-based oscillations | Low–medium ($10^1$–$10^3$) | Time-resolved transcriptomics / proteomics of chromatin remodelers, lamins, condensate proteins; epigenomic states associated with fast/slow relaxation |
| **Poroelastic response / hydraulic permeability**: fluid flow through chromatin–lamina scaffold | Often reported as characteristic drainage time (ms–s); effective permeability or diffusivity (units' model-dependent) | Whole nucleus; can be regionally parameterized in models | ms to s | AFM indentation at multiple loading rates; poroelastic FEM fits; sometimes micropipette aspiration | Low | Joint modelling with transcriptomics of nuclear envelope transport, chromatin density, osmotic regulators; metabolomics of osmolytes and ion handling |
| **Brillouin frequency shift / longitudinal modulus**: optical proxy of high-frequency modulus | GHz shift: relative longitudinal modulus (often reported in arbitrary or normalized units) | 3D submicron maps across entire nucleus in single cells and small tissue volumes | ns to s | Brillouin microscopy (confocal, line-scan, light-sheet variants) | Medium ($10^2$–$10^4$) | Spatial transcriptomics / proteomics in adjacent sections; co-registration with chromatin marks and nuclear bodies (imaging–omics) |
| **Deformability index / transit time** whole-cell or nucleus deformation under flow or constriction | Dimensionless shape index; transit / entry time (ms–s) | Whole cell (often dominated by nucleus in many lineages) | µs to s | Real-time deformability cytometry (RT-DC); microfluidic constriction/filtration assays; aspiration-through-pore devices | High ($10^4$–$10^6$) | Mechanics-indexed single-cell RNA-seq, mutational profiling of deformable vs rigid subclones, epigenetic states linked to mechanotype |
| **Microrheological parameters (G,' G'', MSD exponent α)** local viscoelasticity from fluctuations / driven motion | Storage / loss moduli (Pa) vs frequency; α (0–1) for anomalous diffusion | Subnuclear (100 nm–µm) domains; local chromatin/condensate environments | ms to s | Multiple particle tracking (MPT) of endogenous granules or beads; AFM-based micro rheology; optical tweezers micro rheology | Low–medium (10–$10^3$) | Co-localization with chromatin marks (ChIP-seq / CUT&RUN), ATAC-seq accessibility, LLPS proteomics, single-cell transcriptomics of phase-separation regulators |
| **Local strain field / deformation patterns** spatiotemporal maps of nuclear deformation | Strain (dimensionless), strain rate ($s^{-1}$) | Subnuclear to whole-nucleus maps | s to min | Deformation microscopy (image-based warping), high-resolution fluorescence imaging under defined loading, 3D traction-force microscopy coupled to nuclear markers | Medium ($10^1$–$10^3$) | Spatial transcriptomics and chromatin conformation (Hi-C/HiChIP) in matched geometries; integration with mechanosensitive gene sets and damage/repair signatures |

**Table 2 Instrumental Techniques.** Comparison of the main instrumental techniques used for the mechanical characterization of the cell nucleus and the whole cell. For each technique, the underlying physical principle, the spatial scale of measurement, whether it is contact or non-contact, the typical experimental setup, as well as its most relevant advantages and key limitations, are summarized. The table includes local and global, low- and high-throughput methods, ranging from high-spatial-resolution approaches for intranuclear mapping to population-based techniques for mechanical phenotyping, along with representative literature references.

| Technique | Physical principle | Spatial scale | Contact | Typical preparation | Main advantages | Key limitations | Reference(s) |
|---|---|---|---|---|---|---|---|
| **AFM (Indentation/mapping)** | Mechanical indentation (physical probe) | 10s–100s nm | Yes | Adherent cells/isolated nuclei | High local resolution. Local E maps | Contact → artifacts, modeling (Hertz/poroelastic) required. Low throughput | Haase & Evans 2015. Hobson 2021 |
| **Brillouin microscopy** | Brillouin spectroscopy (scattering) | ~sub-µm voxel (optical) | No | In situ alive cells | Label-free 3D, non-contact. Suitable for intranuclear mapping | Signal dependent on compressibility/density. Interpretation vs. E requires cross-validation | Bevilacqua et al. 2023. Kabakova 2024 |
| **Multiple Particle Tracking (MPT) /microrheology** | Brownian tracking of endogenous or added tracers | 10s–100s nm (local) | No (if internal particles) | In situ or isolated nuclei | Measures intranuclear heterogeneity. Detects diffusion anomalies and $G'(\omega)$, $G''(\omega)$ | It needs tracers or identification of endogenous granules. Analysis requires good algorithms | Herráez-Aguilar et al. 2020 |
| **Micropipette aspiration (MPA)** | Suction and aspiration | Whole nucleus (µm) | Yes (suction) | Isolated nuclei/cells | Gold standard for overall core viscoelasticity | Low intranuclear spatial resolution. Too much work required (HT versions exist) | Lee & Pruitt 2014. Davidson 2015 |
| **Microfluidic constrictions** | Flow and confinement in channels | µm (global) | No (confinement) | In situ alive cells | High throughput. Simulates confined migration. Quantifies intercellular deformability | Global measurement. Correlation with intranuclear requires combined techniques | Davidson 2015. Agrawal 2022 |
| **Deformability cytometry (RT-DC, RT-Viscoelastic DC)** | Flow image + deformability analysis | µm (whole cell) | No | Suspension cells | Extremely high throughput ($10^3$–$10^5$ cells). Population mechanical phenotyping | Sensitive to cytoskeleton + nucleus. Limited intranuclear resolution | Otto 2015. Asghari 2024 |
| **Optical/Magnetic Tweezers** | Point force by optical beams or magnetic fields | nm–µm local | Yes (manipulation) | In situ (microfilaments, anchors) | Precise measurement of pN forces. Information on lamina-chromatin anchors | Intrusive. Low throughput. Requires probes (beads/labels) | Hobson 2021. J Cell Sci reviews |
| **Microharpoon** | Application of force by microfilament | µm local | Yes | Adherent cells | Studies lamina-chromatin connections. Localized strength | Specialized technique. Low throughput | Hobson 2021 |
| **Fluorescence imaging deformability cytometry** | Nucleus image + deformability | µm | No (fluorescence imaging in flow) | Cells in flow (nuclei stained) | Allows correlation with nuclear markers | Requires staining. Limited intranuclear resolution | Muñoz 2023 |

**Table 3 Main mechanical modelling approaches used to study nuclear heterogeneity**. The table summarizes the several types of models and simulations used to describe nuclear mechanics, indicating whether they are based on continuous or discrete approximations, the relevant spatial and temporal scales, and the key parameters governing their behavior. It also highlights the primary use of each approach for analyzing the differential contribution of the nuclear lamina, chromatin, and nuclear fluids, the experimental methods typically used for their validation, and any associated conceptual or practical limitations. Taken together, these models provide a multiscale framework for interpreting experimental data and linking nuclear organization to its emergent mechanical properties.

| Simulation / Model Type | Continuum or Discrete Mechanics | Spatiotemporal Scale | Key Parameters | Main Use for studying Nuclear Heterogeneity | Validation methods | Limitations | References |
|---|---|---|---|---|---|---|---|
| **Continuum FEM models of nucleus (lamina + chromatin as separate materials)** | Continuum mechanics | µm scale spatial; quasi-static to seconds | Elastic/viscoelastic moduli (kPa/sub-kPa), geometry, boundary conditions | Estimation of lamina and nucleoplasmic moduli; analysis of anisotropy; partition of mechanical contributions | AFM indentation, micropipette aspiration, microfluidics | Simplified constitutive laws; limited resolution of chromatin microstructure; geometry idealization | Dahl et al., 2004 ; Dahl et al., 2008 |
| **Poroelastic models of chromatin–lamina scaffold** | Continuum poroelasticity | µm scale; milliseconds–seconds | Permeability, porosity, modulus, viscosity, drainage times | Explaining rate-dependent stiffness; chromatin–fluid interactions; drainage time scales | AFM rate-dependent indentation | Requires permeability / drainage parameters often hard to measure; coupling assumptions; simplified pore structure | Moeendarbary et al., 2013; Wei et al., 2016; Wei et al., 2025 |
| **Hybrid shell–polymer mechanical models (elastic shell + viscoelastic interior)** | Continuum shell + polymer viscoelastic | µm scale; milliseconds–seconds | Shell modulus, polymer relaxation times, chromatin stiffness | Explaining strain stiffening; buckling suppression; coupling of chromatin small-strain vs lamina large-strain response | Micromanipulation, deformation assays | Shell thickness/chromatin viscoelasticity often parametrized from limited data; large-strain behavior simplified | Banigan et al., 2017; Stephens et al., 2017; Wintner et al., 2020 |
| **Chromatin mechanical models incorporating biochemical modulation (lamin A/C levels, chromatin compaction)** | Continuum viscoelastic | µm scale | Lamina stiffness, chromatin viscoelastic parameters | Linking chromatin state or lamina levels to stiffness; mechanical shifts across cell states | Lamin A/C abundance measurements; chromatin modification assays | Phenomenological moduli; mapping biochemical changes to mechanical changes may be indirect | Stephens et al., 2018; Swift et al., 2013 |
| **Phase-field / soft-matter chromatin models (MELON)** | Continuum phase-field | 100 nm–µm; seconds–minutes | Surface tension, viscosity, interaction parameters | Modeling mesoscale chromatin domains, radial organization, emergent stiffness | Super-resolution imaging; domain segregation patterns | Requires assumptions on phase interactions; limited direct mechanical validation; coarse-grained | Laghmach et al., 2019; Laghmach et al., 2024 |
| **Coarse-grained polymer models (e.g., 1CPN)** | Discrete mechanics (polymer physics) | nm–100 nm; microseconds–seconds | Interaction strengths, persistence length | Linking nucleosome interactions to local flexibility; providing effective mesoscale elastic parameters | Nucleosome-scale structural data | Cannot reach whole-nucleus scale; computationally expensive; coarse-graining choices affect outcomes | Lequieu et al., 2019 |
| **LLPS-based mechanical models of chromatin and condensates** | Continuum or coarse-grained soft matter | 100 nm–µm. ms–min | Interfacial tension, elastic modulus | Describing mechanical microenvironments from phase separation | Condensate imaging, fusion events | Mechanical properties of condensates are often difficult to measure; interfaces idealized | Hyman et al., 2014; Li et al., 2023; Li et al., 2024 |

| Simulation / Model Type | Continuum or Discrete Mechanics | Spatiotemporal Scale | Key Parameters | Main Use for studying Nuclear Heterogeneity | Validation methods | Limitations | References |
|---|---|---|---|---|---|---|---|
| **Multimodal validation frameworks for nuclear mechanical models** | Not a model per se | Technique-dependent | Modulus, viscosity, chromatin organization metrics | Cross-checking FEM/poroelastic predictions; constraining parameters | AFM, Brillouin, deformation microscopy, imaging, Hi-C | Emphasizes parameter degeneracy; relies on multiple modalities | Hobson et al., 2020; Dahl et al., 2004; Kabakova et al., 2024; Wei et al., 2025; Ghosh et al., 2019; Laghmach et al., 2024; Dupont & Wickström, 2022 |